# Bibliometric Study of Authorship Pattern Literature

Deep Kumar Kirtania

Librarian, Bankura Sammilani College

deepkrlis@gmail.com

**Abstract:** The main objective of this study is to conduct a bibliometric analysis of scholarly publications of Authorship Pattern. The present study covers 1723 research papers published in the area of authorship pattern and indexed in Scopus database from the year 2013 to 2022. These research publications considered for the present study have been analysed based on their year wise growth, pattern of authorship, times citations, type of publication, most productive publication source as well as countries and institutions. The study shows the positive growth of the literatures with collaborative authorship pattern and good citation status. This original research paper will be helpful to the researchers of library and information science, especially who are working in the area of bibliometrics studies.

**Keywords:** Bibliometric Analysis, Authorship Pattern, Authorship Study, Collaborative Authorship, Degree of Collaboration, Scopus.

**Introduction:** Authorship study is a type of bibliometric analysis that focuses on examining the patterns and characteristics of authorship in scholarly publications (Khaparde & Pawar, 2013). This includes analysing the number of authors per publication, author affiliations, author collaborations, author productivity, and author impact (Mahapatra, 2002). In authorship studies, bibliometric techniques are used to analyse large datasets of scholarly publications to identify patterns and trends related to authorship (Bandyopadhyay, 2001). In recent years, most of the research has been conducted jointly for the overall growth of information communication technology worldwide (Patra, Bhattacharya & Verma, 2006; Kirtania, 2021). This collaboration pattern may have one to multiple authors, and may involve domestic or international researchers from any organization in the country, starting with colleagues from the same organization (Kirtania & Chakrabarti, 2018). In recent times, there has been a lot of work on authorship patterns, but the current status of existing publications on authorship pattern, growth, times cited, core publication source should be analyzed. This is done in the present paper, with the help of bibliometric analysis, as bibliometric analysis is a quantitative research method that involves the use of statistical and computational techniques to analyze the characteristics and patterns of scholarly publications (Patra, Bhattacharya & Verma, 2006).

**Objectives:** The objectives of this study are:

- to examine the year wise growth of the publications. and their citations.
- to trace out the authorship pattern of these publications and
- to analyse the citation count, type of publications, most productive affiliated institutes and country of the publications.

**Methodology:**

**Scope & Coverage:** This study covers 1723 research publications on "authorship pattern" and indexed in Scopus database from 2013 to 2022.

**Method Used:** First, the Scopus database was searched for publications using the term "authorship pattern," and then the papers from the ten-year period of 2013 to 2022 were retrieved by applying data filtering to the results. As a result of the query and filtering, 1723 papers on authorship pattern were found and selected for the present study. Each publication was then assessed for bibliographic data collection, such as year of publication, number of authors, affiliated country and institute, times cited, and type of publication. The raw data were collected, stored, organized, and presented separately in MS-Excel. The data were then tabulated, analysed and interpreted to draw final conclusions.

**Data Analysis and Findings**

**Table1: Year wise distribution of publications**

| Year | Documents | Percentage | % Change |
|------|-----------|------------|----------|
| 2013 | 114 | 6.62 | N/A |
| 2014 | 118 | 6.85 | 3.39 |
| 2015 | 129 | 7.49 | 8.53 |
| 2016 | 158 | 9.17 | 18.35 |
| 2017 | 139 | 8.07 | -13.67 |
| 2018 | 156 | 9.05 | 10.90 |
| 2019 | 197 | 11.43 | 20.81 |
| 2020 | 212 | 12.30 | 7.08 |
| 2021 | 310 | 17.99 | 31.61 |
| 2022 | 190 | 11.03 | -63.16 |
| **Total** | **1723** | **100.00** | **N/A** |

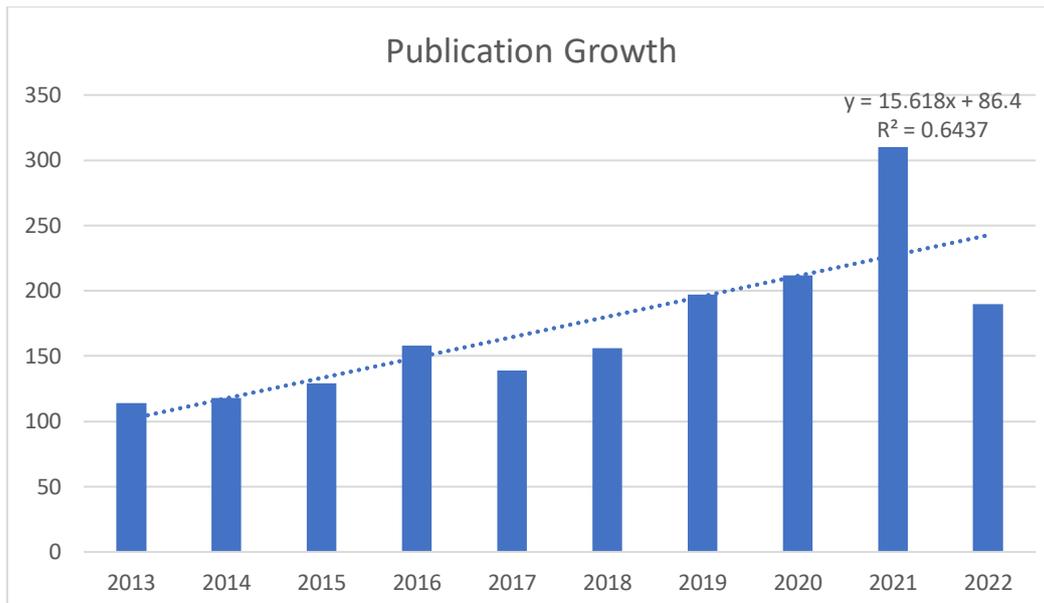

**Figure 1: Publication Growth of the Authorship Pattern**

**Literature Growth:** Table 1 describes the year-wise distribution of publications related to authorship patterns. It can be seen that a total of 1723 research papers have been published on authorship patterns in the ten years from 2013 to 2022, which is statistically significant. Looking at the year-on-year publication growth (Figure 1), it is easy to see a positive trend (y = 15.618x + 86.4). In terms of individual year-wise publications, 310 articles were published in 2021, followed by 2020, 2019, and 2022, respectively. Although the publication trend is positive, the number of research papers published in 2022 has decreased slightly, which does not follow this trend line.

**Table 2: Authorship pattern of the papers**

| Number of Author | Number of Papers | Percentage |
|---|---|---|
| One | 323 | 18.75 |
| Two | 520 | 30.18 |
| Three | 375 | 21.76 |
| Four | 239 | 13.87 |
| Five | 125 | 7.25 |
| More than five | 141 | 8.18 |
| **Total** | **1723** | **100.00** |

**Authorship Patten & Degree of Collaboration:** Table 2 describes the authorship pattern of the publications. It can be seen that only 18.75% of all published articles were written by single authors, while the majority were written by joint or collaborative authors. Among the collaborative research works, the number of articles published by two authors is the highest at 30.18%, followed by three authors (21.76%) and four authors (13.87%). The degree of collaboration indicates the trend of collaborative authorship pattern among the authors for publishing papers (Subramanyam, 1983). The average degree of collaboration is 0.81, which clearly indicates a slight dominance of joint authors in the field of authorship pattern."

Degree of Collaboration is calculated by simple formula that is (DC) = $\frac{Nm}{Nm+Ns}$ where Nm is the number of multi-authored papers, and Ns is number of single authored papers (Subramanyam, 1983).

Table 3: Time Cited of the publications

| Time Cited | Number of Papers | Percentage |
|---|---|---|
| 0 | 459 | 26.64 |
| 1 | 226 | 13.12 |
| 2 | 148 | 8.59 |
| 3 | 108 | 6.27 |
| 4 | 87 | 5.05 |
| 5 | 73 | 4.24 |
| 6 | 69 | 4.00 |
| 7 | 49 | 2.84 |
| 8 | 41 | 2.38 |
| 9 | 43 | 2.50 |
| 10 | 29 | 1.68 |
| >10 | 391 | 22.69 |
| **Total** | **1723** | **100.00** |

**Citation Status:** Table 3 presents the citation count for the publications, and it shows that three-quarters of the 1723 papers have received at least one citation, which is quite impressive. Furthermore, 420 articles have been cited 10 or more times, representing about 24% of all publications, which is a significant statistic in terms of citations. Analyzing the citation behavior from this table, it can be concluded that the articles on this topic are of high quality, which is why they have received a considerable number of citations

**Publication Source:** Table 4 shows the top ten sources of publication, and it can be seen that papers on authorship patterns are published in reputed journals all over the world, indicating that these papers are of very good quality. Among the publication sources, Library Philosophy and Practice occupied the first place with 21.71% of the total publications, followed by Scientometrics (82), Lecture Notes in Computer Science, DESIDOC Journal of Library and Information Technology, and others

**Table 4: Top Ten Source of Publication**

| Publication Source | Number of Papers | Percentage | Rank |
|---|---|---|---|
| Library Philosophy and Practice | 374 | 21.71 | 1 |
| Scientometrics | 82 | 4.76 | 2 |
| Lecture Notes in Computer Science | 45 | 2.61 | 3 |
| DESIDOC Journal of Library and Information Technology | 34 | 1.97 | 4 |
| Plos One | 31 | 1.80 | 5 |
| Journal Of Informetrics | 18 | 1.04 | 6 |
| Annals Of Library and Information Studies | 15 | 0.87 | 7 |
| Current Science | 12 | 0.70 | 8 |
| Ceur Workshop Proceedings | 9 | 0.52 | 9 |
| International Journal of Information Science and Management | 9 | 0.52 | 9 |
| Science And Technology Libraries | 8 | 0.46 | 10 |

**Table 5: Most productive countries**

| Country | Number of Papers | Percentage |
|---|---|---|
| India | 504 | 29.25 |
| United States | 302 | 17.53 |
| United Kingdom | 98 | 5.69 |
| China | 95 | 5.51 |
| Spain | 95 | 5.51 |
| Canada | 58 | 3.37 |
| Brazil | 56 | 3.25 |
| Saudi Arabia | 56 | 3.25 |
| Germany | 55 | 3.19 |
| Pakistan | 55 | 3.19 |

**Most productive countries:** Table 6 describes of the publications by country shows that India tops the list with 504 articles published, followed by the United States (302) and the United Kingdom (98). In addition, countries such as China, Spain, Canada, Brazil, Saudi Arabia, Germany, and Pakistan have also made significant contributions to the research on authorship patterns

**Table 6: Top Ten Most Productive Institution**

| Institution Name | Number of Papers | Percentage | Rank |
|---|---|---|---|
| Periyar University | 30 | 1.74 | 1 |
| Imam Abdulrahman Bin Faisal University | 29 | 1.68 | 2 |
| Alagappa University | 24 | 1.39 | 3 |
| Banaras Hindu University | 22 | 1.28 | 4 |
| Universitat de València | 19 | 1.10 | 5 |
| Annamalai University | 17 | 0.99 | 6 |
| King Saud bin Abdulaziz University for Health Sciences | 17 | 0.99 | 6 |
| Universidade de São Paulo | 16 | 0.93 | 7 |
| University of Kashmir | 15 | 0.87 | 8 |
| University of Delhi | 15 | 0.87 | 8 |
| Kalinga Institute of Industrial Technology, Bhubaneswar | 14 | 0.81 | 9 |
| Prince Sultan University | 13 | 0.75 | |
| Chinese Academy of Sciences | 13 | 0.75 | |
| Mizoram University | 13 | 0.75 | |

**Productive Institutions:** An analysis of the institutional affiliations (Table 6) of the authors shows that prestigious educational institutions from around the world are associated with these research publications. Periyar University has occupied the number one position in the list by publishing 30 articles, followed by other reputed institutes such as Imam Abdulrahman Bin Faisal University (29), Alagappa University (24), and Banaras Hindu University (22), among others

**Table 7: Type of Publication**

| Type of publication | Number of Papers | Percentage |
|---|---|---|
| Article | 1315 | 76.32 |
| Conference Paper | 205 | 11.90 |
| Review | 111 | 6.44 |
| Book Chapter | 35 | 2.03 |
| Conference Review | 21 | 1.22 |
| Erratum | 15 | 0.87 |
| Book | 8 | 0.46 |
| Note | 6 | 0.35 |
| Letter | 3 | 0.17 |
| Short Survey | 2 | 0.12 |
| Editorial | 1 | 0.06 |
| Retracted | 1 | 0.06 |
| **Total** | **1723** | **100** |

**Type of Publication**: From the above table it can be seen that out of the total papers, research articles are the most published type of publication, thereafter conference paper, review, book chapter and conference review respectively have been published. A variety of topics have been published on authorship patterns, which can be considered a very significant finding of this present study.

**Conclusion:** Currently, researchers are working on various topics ranging from domestic to national and international levels. In this context, the aim of authorship studies is to analyse all the parameters associated with publications in any research field. Authorship studies have been conducted on various subjects, from science to humanities and social sciences in the past. Analysing the current state of research publications on authorship studies was essential and the main objective of this work. This study shows that the publication growth trend of authorship patterns has been very positive, with joint authorship patterns found in these publications, and high-quality sources of publication. Additionally, the citations of the papers indicate that publications on this topic are highly cited, which is significant. Analysing these parameters suggests that more quantitative research should be conducted in the future by applying authorship studies to other disciplines.